\documentclass[submission,copyright,creativecommons]{eptcs}
\providecommand{\event}{WS-FMDS 2012} % Name of the event you are submitting to
\usepackage{breakurl}             % Not needed if you use pdflatex only.
\input pdfcolor.tex 
\usepackage{amssymb}
\usepackage{amsmath}
\usepackage[pdftex,dvips]{graphicx}
\usepackage[usenames,dvipsnames]{color}
\usepackage{textcomp} % \textlbrackdbl, \textrbrackdbl
\usepackage[cp1252]{inputenc}
\usepackage{xspace}
\newcommand{\Focus}{\textsc{Focus}\xspace}
\newcommand{\isah}{Isabelle/HOL\xspace}
\newcommand{\isabh}{Isabelle/HOL\xspace}

\newcommand{\focus}{\Focus}
\newcommand{\autofocusString}{\textsc{AutoFocus}}
\newcommand{\autofocus}{\autofocusString\xspace}
\newcommand{\af}{\autofocus}
\newcommand{\autofocusIII}{\autofocus~3\xspace}
\newcommand{\afIII}{\autofocusIII}
\newcommand{\aft}{\autofocusIII}
\title{Verified System Development\\ with the AutoFocus Tool Chain}
\def\titlerunning{Verified System Development}

\author{Maria Spichkova
\institute{Institut für Informatik, TU München,\\ 
Boltzmannstr. 3, 85748 Garching, Germany}
\email{spichkov@in.tum.de}
\and
Florian Hölzl \qquad David Trachtenherz
\institute{fortiss GmbH, \\
Guerickestr. 25, 80805 München, Germany}
\email{\{hoelzlf,trachten\}@fortiss.org}
}

\def\authorrunning{M. Spichkova, F. Hölzl, D. Trachtenherz}
\begin{document}
\maketitle

\begin{abstract}
This work presents a model-based development methodology\footnote{This work was fully funded by the German Federal Ministry
of Education, Science, Research and Technology (BMBF) in
the framework of the Verisoft XT project. The responsibility for this article lies with the authors.} 
for verified software systems as well as a tool support for it: 
an applied \aft{} tool chain and its basic principles emphasizing the verification of the system under development as well as the check mechanisms we used to raise the level of confidence in the correctness of the implementation of the automatic generators.
\end{abstract}

\section{Introduction}
\label{sec:intro}

Software-based system development has become one of the most challenging fields of software engineering research and industrial application.
To support the contemporary system development in industry CASE tools are used 
-- they allow a simple and (mostly) intuitional design of distributed systems and applications, and
executable code is generated directly from the models developed using these tools.
In state-of-the-art industrial development, quality assurance is performed by extensive testing of the generated code. 
However, testing can only demonstrate the absence of errors for exemplary test cases, 
but not the correctness of the system. In opposite to testing, 
formal verification delivers a correctness proof for safety critical properties but requires significant effort. 
Thus, only the most critical parts of a system must be verified, 
and the whole process of specification and verification must be set up in a way that minimizes the overall effort -- 
the verification process must be integrated into model-based development of safety-critical systems.  
On the other hand, testing and simulation must belong to the development process, because in some cases, even after verifying certain properties,
inconsistencies can still remain in the specification, model or code -- 
most often an important property is overlooked as nicely stated by Donald E. Knuth's famous saying: 
``Beware of bugs in the above code -- I have only proved it correct, not tried it.''

There is a number of works on integration of different system models 
and verification techniques. 
For instance, the Ptolemy approach (see \cite{ptolemy:ieee2003}) introduces a general way to combine heterogeneous models of embedded systems. A prominent example of integration of verification techniques is a combination of Model Checking and Deduction for I/O-Automata done by O.\  Müller and T.\ Nipkow (see \cite{MullerNipkow:TACAS1995}).
However, to our best knowledge there are no other works 
on achieving a \emph{pervasive formal development process} for embedded applications 
starting with informal textual specification and leading to verified machine-code.
This direction has been touched for the first time in \cite{verisoft_journal},
though only for upper layer of automotive systems and focused on later verification phases -- 
in contrast, the contribution of the work presented here is covering the entire seamless pervasive development process.

The first steps towards a methodology for development of verified embedded system have been done in~\cite{VerisoftAutomotive_FM06,BoKoKuSp05}.  
For example, a typical setting found in the automotive domain, a time-triggered operating and communication bus system, has been verified \cite{efts_book,verisoft_journal}.
In this paper we deal with the verification of the application software. 
In comparison to the problem frame approach of M. Jackson~\cite{jackson} as well as the 4-variable model of Parnas and Madey~\cite{parnas_madey},  we present a pervasive formal development methodology for embedded systems 
starting from an informal textual specification of the requirements and going all the way to verified application code.
Earlier results of the Verisoft project~\cite{verisoft_journal} have shown the methodology for later verification phases,
in particular the relation between the application model and its execution environment, e.g. the operating system. 

We outline here a development methodology for safety-critical systems with focus  on the tool chain, which is employed in the design, implementation and verification phases of the methodology. 
The tool chain emphasizes verification of the system under development -- it not only allows integration into the process of modeling, but also reduces the verification and integration effort.

\section{Development Methodology}
\label{sec:methodology}

\begin{figure}
	\centering
		\includegraphics[scale=0.55]{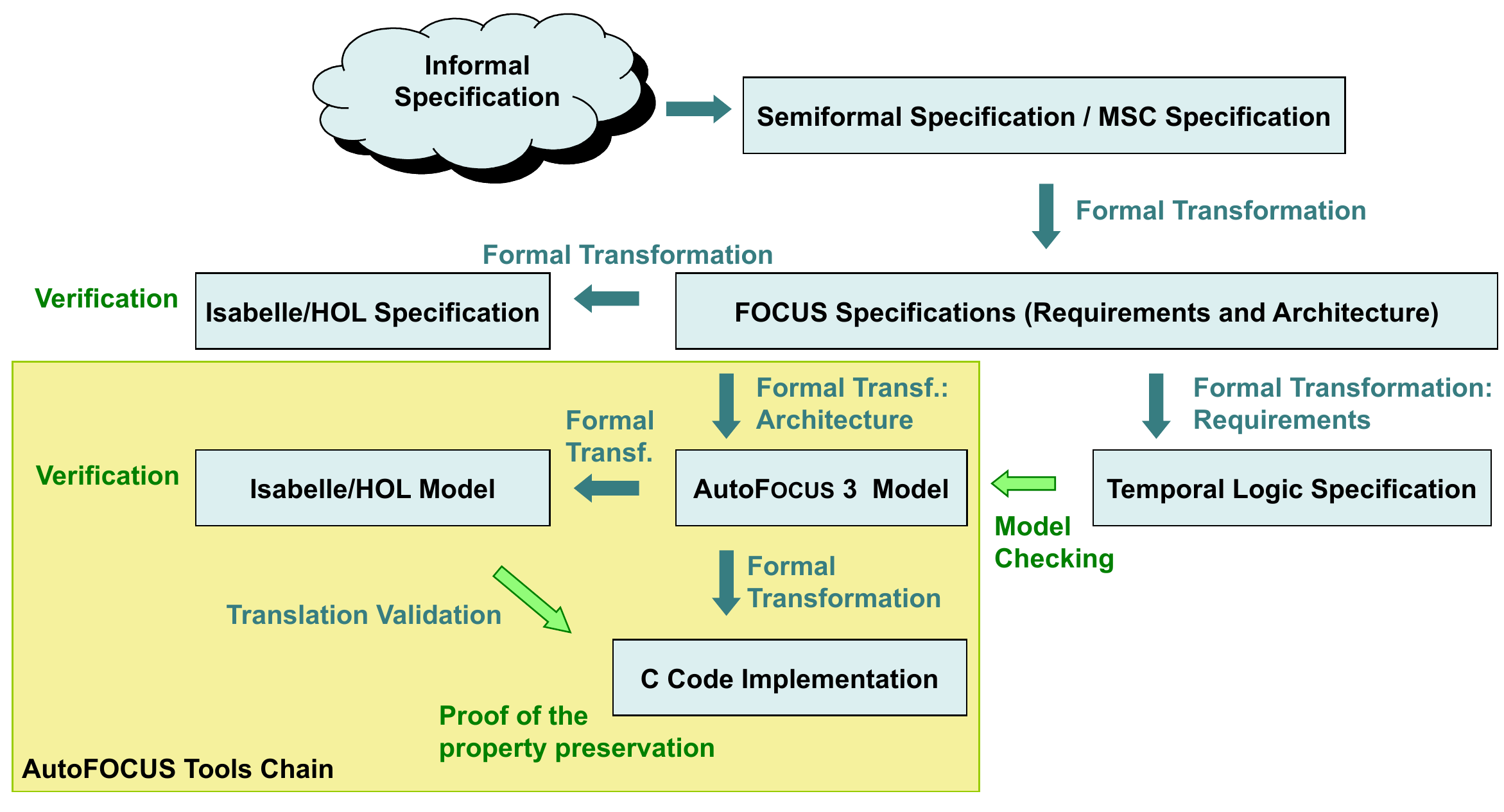}
	\caption{Development Methodology}
	\label{fig:methodology}
\end{figure}

Fig.~\ref{fig:methodology} illustrates the structure of the development methodology in a top-down manner: from an informal specification through multiple transformation steps we get a verified formal specification, a verified executable model and also a verified C code 
implementation. The boxes represent development artifacts, the  dark arrows show the dependency relations, i.e.,~which artifact is used as input for the development of the successor artifact. The light arrows show the proof relations between the artifacts.

Beginning with a \emph{requirements specification} (we focus here only on functional requirements, including timing aspects), which captures the relevant aspects of the system to be developed in flexible yet informal way, we derive  a \emph{tabular semi-formal specification}.
This first step raises the level of precision by transforming the free text requirements into a structured form using specific pre-defined syntactic patterns as presented in \cite{fleischmann08}. 
An informal specification consists of a set of words, which can be distinguished into two categories: content words and keywords (relation words).
Content words are system-specific words or phrases, e.g. \emph{``system is initialized''} or \emph{``Off-button is pressed''}. The set of all content words forms the \emph{logical interface} of the system, which can be understood as some kind of (domain specific, system-dependent) glossary that must be defined in addition. 
Keywords are domain-independent and  form relationships between the content words (e.g. ``if'', ``then'', ``else''). 
A semiformal specification consists of a number of requirements described using the following textual pattern, which is easily be understood also by engineers unfamiliar with the formal methods:%
\[
\begin{array}{ll}
WHILE & {Some~state}
\\
IF    & {Some~event~occurs~or~some~state~changes}
\\
THEN  & {Some~event~occurs~or~some~state~changes}
\\
ELSE  & {Some~event~occurs~or~some~state~changes}
\end{array}
\]
An \emph{event} describes a \emph{point in time}, in which the system observes or does something; the duration of the event is not important, e.g., ``driver presses a button''. A \emph{state} describes a system or component state \emph{within some time period}, e.g., ``a button is pressed''.  
Using such a simple description to structure the information from the informal specification, we can find out missing information quite fast. 
Furthermore, we identify possible synonyms that must be unified before
proceeding to a formal specification. Analysis of the semiformal specification document should also yield sentences, which need to be reformulated or extended.
This specification can be also rewritten to a \emph{Message Sequence Charts} (MSCs, cf.~\cite{MSC:Kluwer2003}) 
representation according to the approach presented in~\cite{dentum_focus_tb}. 
The purpose of using MSCs  for specification of  highly interacting systems  is to obtain a better overview in comparison to a textual representation. 

Subsequently, the specifications are translated to \Focus~\cite{focus}, a framework for formal specifications
and development of distributed interactive systems. 
\Focus is preferred here over other specification frameworks since it 
has an integrated notion of time and modeling techniques for unbounded networks  
(where we have replications of system components of the same kind), 
provides a number of specification techniques for distributed systems and concepts of refinement, as well as graphical notation, which is extremely important when we are dealing with systems of industrial size. 
In general we represent in  \Focus  two kinds of specifications: 
a formal specification of \emph{system requirements} and the corresponding \emph{architecture specification}. 
Both of them are extracted from the MSC specification and/or the semiformal specification. 
This representation prepares the ground to 
verify the system architecture  specifications against the system requirements by translating both 
to the theorem prover Isabelle/HOL~\cite{npw} via 
the framework ``\Focus on Isabelle''~\cite{spichkova}.

As the next step, we translate the architecture specification 
to a representation in the related CASE tool
{\afIII}~\cite{Schaetz:Kluwer2004,af3homepage} to simulate the system. 
The requirements specification can be translated from \Focus to temporal logic, which gives basis to model-check the model.
The transformations from \Focus to temporal logic and to the {\afIII} representation are formal and schematic, given some constraints on the \Focus specification are obeyed.
This model can also be exported to Isabelle/HOL to prove its properties -- it is in general a \emph{refinement} of a \Focus specification, thus its properties can be slightly different, i.e., more strict, from the ones specified on the \Focus layer, but the proof schema, which has been developed for the \Focus specifications, can be (partially) reused. 
 
The last step in our methodology is transformation of the model to a corresponding C code by a code generator: we show that the C program is an admissible simulation
of the \aft model. 
This concludes our methodology, but that is
of course not the end of the complete development process: we need to deploy the code into the execution
environment. %
Altogether, the methodology guides us from an informal specification via stepwise refinement to a verified formal specification, a corresponding executable verified model, and also a corresponding verified C code implementation.

The applicability of the methodology has been demonstrated by two case studies from automotive area. 
The first case study~\cite{dentum_tb,dentum_focus_tb}, developing an Adaptive Cruise Control (ACC) system with Pre-Crash Safety functionality,  was motivated and supported by DENSO Corporation, the second  case study \cite{spichkova_tb_decomp}, developing a Cruise Control System with focus on system architecture and system verification, was supported by Robert Bosch GmbH, %  
Table~1 gives short description how the statistics about these two case studies, 
the size of the \aft model is approximately equal  to that of the corresponding \focus specification.

 In each of the case studies we had a final number of system states was final according to the \aft modeling paradigm, the most complex component was the component specified the main logic of the system. In the case of the Cruise Control System we split the specification of the main system logic into four components according to the formal decomposition approach~\cite{spichkova_tb_decomp} introduced within the system development methodology.  
There is a large number of approaches in area of decomposition  (see, e.g., \cite{Hofmann_approachesto,PR99,2006-01-1222}). 
The main difference and the main contribution of our decomposition methodology is that it  was developed for such a system architecture, 
where we already know (or, more precisely, have already specified them in a formal way) systems or components properties and need to decompose this whole properties collection to a number of subcomponents to get readable and manageable specifications.

An sample-property proven for  the Cruise Control System looks like follows: 
\emph{If the driver pushes the Accelerate-button while the system is on and none of the  switch-off constraints occurs, the system must accelerate the vehicle during the next time unit respectively to the current speed and the predefined acceleration schema.} 
This means also that the system must analyze the information from sensors to check whether any switch-off constraints occurs, i.e. if the battery voltage is too low or if the gas pedal sensor fails. 

\begin{table}[ht]
\begin{center}
\begin{tabular}{|l|l|l|}
\hline
Methodology Artifact & ACC / Pre-Crash Safety & Cruise Control: Architecture 
\\
\hline
\hline
Semiformal Specification (requirements) 	&  30  &  	70 
\\
MSC Specification  (MSCs) 				&  		 16     		& 	 20 
\\
FOCUS Specification  (components)  			& 	10 	& ca. 80
\\
FOCUS Specification (hierarchy levels) 		&  2  		& 4
\\
Isabelle Specification (lines of code)		&  -  		& ca. 12,500 
\\
Isabelle Model (lines of code)			&  -  		& ca. 38,000  (generated)
\\
Verification of Isabelle  Specification	&  -  		& ca. 8,500 lines of proof
\\
Partial Verification of Isabelle Model	&  -  		& ca. 5,000 lines of proof
\\
C-Code ( lines of generated code)			&  ca. 3,000  & ca. 17,000  \\
\hline
\end{tabular}
\caption{Case Studies Statistics}
\end{center}
\label{tab:cs}
\end{table}%
\section{AutoFOCUS}
\label{sec:af}

\aft  \cite{af3homepage,Schaetz:Kluwer2004} is a scientific prototype implementing a modeling language
based on a graphical notation   designed to support the modeling of distributed, timed, reactive systems.  
 The tool a restricted version of the formal \focus semantics, in particular the
time-synchronous frame and supports differents layers of models, their simulation and analysis. 
Main aspects of \aft are
hierarchical component decomposition of the system,
modeling behavior of application parts of the system as well as
modeling of technical architectures including control units and communication busses.

We give a brief
introduction of the current language. 
In particular, we introduce three modeling views: \emph{data type},
\emph{system structure}, and \emph{component behavior} specifications.

Every computer-based system can be seen as a data processing system. Thus programming and modeling languages
are based on some mechanism for specifying data items or more general data types.
\aft uses a functional language to specify data types, this provides a set of predefined
types, such as Boolean and Integer. Based on these, we can define more complex types using variant
or enumeration types and record-like types. User-defined functions allow us to implement additional operations
for our data types.

Although in general data types and user functions may be recursive, we restrict them to be non-recursive
for the domain of embedded systems, because this eases several tasks.  Firstly, we obtain statically computable
bounds for memory consumption and execution times. For embedded software this is often highly desirable,
because of restricted hardware resources and the need to respect given timing bounds and constraints.
Secondly, current model-checkers rely on bounded types.

The \emph{system structure specification} is similar to the \focus architecture specification. It captures
the static aspects of the system description. We specify a network of communicating components
working in parallel (assuming a global synchronized time frame). Each component has a syntactic
interface described by a set of ports. Each port is either an input or an output port, has a data type
and an initial value. 
Furthermore, each component is declared to be weakly causal or strongly causal.
Weak causality models instantaneous reaction, while strong causality models a delayed
reaction.
The network of components is formed by connecting ports with channels. From the semantics
point of view, we need to avoid weakly causal feedback loops (Brock-Ackermann anomaly).
Fortunately, this can be checked easily by the tool's model constraint checker.

System structure specifications may be separated into hierarchic
views n order to deal with larger models. 
Components can be refined into a set of sub-components introducing both local communication
and communication to the environment, e.g., through the interface of the parent component.

Atomic components have their behavior specified using one the following variants: a stateful
\emph{automaton specification} or a stateless \emph{function specification}.
An automaton specification describes an input/output automaton. It consists of a set of
control states, a set of typed data state variables, and a set of transitions. One of the control
states is marked as the initial state. Every data state variable is also initialized to a given
value.

Each transition has a source and a target control state. Furthermore, each transition
specifies patterns of messages received via the input ports of the respective component
and preconditions over the inputs and the data state variables. A transition can fire, if its
source state is the current control state, the current input values match its input patterns,
and the precondition is fulfilled. If a transition fires, the current state of the component changes to its target state, the output
values of the component are updated according to the output pattern specification of the
transition, and the data state variables are also updated as specified by the postcondition
part of the transition specification. 
Note that more than one transition might be enabled for a given component state and set of
input port values. Thus, component behavior can be non-deterministic.

Sometimes it is cumbersome to give an automaton specification for simple components. In
particular, if the component does not need an internal state, a simple mapping from input
values to output values is convenient. In \aft this is modeled by a function specification.
This is a simple table where each line defines such a mapping from inputs to outputs. Again,
the table can define non-deterministic choices. From the semantic point of view function
specifications can be seen as restricted automaton specifications.

Using the three views introduced above, we obtain an \emph{executable model}. We can now validate
the model using the \aft simulator to get a first impression of the system under development
and possibly find implementation errors that we introduced during the manual transformation
of the \focus specification into a \aft model. Automatisation of this transformation is an ongoing work.

\section{Generation Tool Chain}
\label{sec:tools}

Our goal is to establish a sound verification environment with a high confidence in the correctness of the environment and thus of correctness the designed system, for this purpose  
we have combined \aft{} and the theorem prover environment \isabh by implementing 
a generator for the formal transformation of the design model into the theorem prover model, as well as 
a code generator for producing a C code implementation of the design model.
 
The central part is the transformation of the design model into the C code, i.e., the implementation path of the product under development.
This is supported by two verification mechanisms: semi-automatic verification with a theorem prover and fully automatic verification by means of a model-checker.
These flanking measurements fulfill two purposes.
On the one hand they verify actual properties of the design model and the implementation code,
On the other hand differing results of both techniques indicate an implementation fault in one of the generators.

When building a tool chain based on automatic generators it is vital to take great care about what one is doing -- 
one must understand the semantics of the generation input, e.g., the \aft{} model's semantics, and the target language, e.g., the C language.
Since automatic code generation only makes sense, if the behavior of the generated program is equivalent to the behavior of the input model, we must show that the transformation is preserving the semantics. 
We strongly emphasize that the semantics of the design language has been defined before implementing the generators.
Unfortunately, many code generation environments do not follow this up-front approach.
They either do not care about behavior at all (i.e., generate structural outputs only)
or consider the semantics of the generation input to be defined by the semantics of the generation outputs (i.e., the semantics of the design model is defined by the semantics of the generated code).
The first case is of no further interest to us, since the generator only produces hull code that needs to be extended manually.
The second case forces the user of the input language to work around flaws of the generator as we have seen with bugs in compilers.
Usually, such flaws are exploited by flaws in the developed system, while in our approach the semantics of the design language provides the basis for the specification of the generators.

\subsection{The \isah Translator}
\label{sec:isa-gen}

Formally specified requirements can be verified using a theorem prover: 
for this purpose we have developed a translator,
which generates \isah theories from \af models.
The behavioural equivalence between \af models
and their generated representation in \isah
has been shown in a paper-and-pencil proof \cite{isabelleExport}. 

The \isah code for an \af model is created as follows.
Firstly, the user initiates code export for the data dictionary,
which generates a theory containing data type declarations and function definitions used in the model. 
Then the user may generate code for any of the components in the model: 
when selecting an atomic component, a theory will be generated
containing the input/output interface definition for the component
and the transition function originating from the automaton
or function specification used to define the component's behaviour;
when selecting a composite component,
recursively the theories for all subcomponents will be generated and,
ultimately, the theory for the considered component,
whose transition function 
performs data transmission between the subcomponents
and invokes all subcomponents' transition functions.
Generating code for the root component of an \af model
yields a set of theory files encoding this model in \isah, as well as 
proof of theorems 
that support subsequent verification of model's properties
in \isah.  
\subsection{The C0-Code Generator}
\label{sec:c-gen}

In order to obtain executable code from an \aft model, we have 
implemented a C0 code generator.
A paper and pencil proof (cf.~\cite{c-code-gen}) has shown the behavioral equivalence between the model and the generated code. 

C0 is a C language subset constructed for usage with the \isah verification
environment as discussed in \cite{Schirmer:PhD}. 
It  differs from C by restricting the language -- a number of hazardous features, like pointer arithmetic, are
forbidden in C0, which eases the reasoning
and verification with \isah. 
As a result of these restrictions, we gain the advantage of being able to compute memory consumption at compile time
as well as worst case execution times automatically, since all operations and function calls are non-recursive. 
\cite{Petrova:phD} and \cite{Leinenbach:PhD}
present the verification of a non-optimizing C0 compiler, which was itself written in C0 with the tool \emph{aiT} by AbsInt \cite{absint}.

\section{Conclusions and Future Work}
\label{sec:future}

We have presented  the model-based development methodology for verified software systems as well as the corresponding \aft{} tool chain to explain how this tool chain and its principles can be applied for verification. Our focus was on the design, implementation and the verification phase of the overall methodology.
The feasibility of the proposed approach   
has been demonstrated by two  industrial case studies from automotive area:
one case study  was motivated and supported by DENSO Corporation, the second case study was supported by Robert Bosch GmbH. 

The formal transformations from \Focus to Isabelle/HOL as well as to \aft  
are specified in the presented methodology as schematic, but not as automatic ones.
Automatization of these steps is an ongoing work. 
We are currently also  complementing the \aft language with a description language for distributed execution
environments based on embedded control units and bus systems. 
Furthermore, a deployment model combines the logical architecture with the technical architecture,
thus resulting in a complete description of the executable system. We have to extend
our methodology into this area by giving suitable verification methods on these new models,
as done for a time-triggered operating and bus system in \cite{efts_book}.

\bibliographystyle{eptcs}
\bibliography{biblio2}

\end{document}